\documentclass[12pt]{article}
\usepackage{amssymb,amsmath,epsfig}

\begin{document}

\title{\bf Reconstruction of Scalar Field Dark Energy Models in Kaluza-Klein Universe}
\author{M. Sharif \thanks {msharif.math@pu.edu.pk} and Abdul Jawad\thanks {jawadab181@yahoo.com}\\
Department of Mathematics, University of the Punjab,\\
Quaid-e-Azam Campus, Lahore-54590, Pakistan.}

\date{}

\maketitle
\begin{abstract}
This paper is devoted to study the modified holographic dark energy
model by taking its different aspects in the flat Kaluza-Klein
universe. We construct the equation of state parameter which
evolutes the universe from quintessence region towards the vacuum.
It is found that the modified holographic model exhibits instability
against small perturbations in the early epoch of the universe but
becomes stable in the later times. We also develop its
correspondence with some scalar field dark energy models. It is
interesting to mention here that all the results are consistent with
the present observations.
\end{abstract}
\textbf{Keywords:} Kaluza-Klein cosmology; Modified holographic
dark energy; Scalar field models.\\
\textbf{PACS:} 04.50.Cd; 95.36.+x.

\section{Introduction}

During the last decades, astrophysicists and astronomers have made
remarkable predictions about the gap of more than $70$ percent of
the overall energy density in the universe. Based on recent
observations \cite{1,2}, it has been made consensus that dark energy
(DE) fills this gap. It is the unknown force having large negative
pressure and referred to be responsible for accelerated expansion of
the universe. Many efforts have been made about its identity but its
nature is still unknown. The most obvious candidate of DE is the
cosmological constant (or vacuum energy) but it suffers two major
problems \cite{2a}. It is an interesting as well as the most
challenging problem to find the best fit model of DE.

There are different dynamical DE models out of which holographic
dark energy (HDE) model is the most prominent. It is proposed in the
scenario of quantum gravity on the basis of holographic principle
\cite{3}. In fact, the derivation of this model is based on the
argument \cite{4} that \textit{the vacuum energy (or the quantum
zero-point energy) of a system with size $L$ should always remain
less than the mass of a black hole with the same size due to the
formation of black hole in quantum field theory}. In mathematical
form, we have $\rho_{\Lambda}=\frac{3c^{2}}{8\pi GL^{2}}$, known as
HDE density \cite{5,6}. Here the constant $3c^2$ is used for
convenience, $L$ represents the infrared (IR) cutoff and $G$ is the
gravitational constant.

The choice of IR cutoff in the HDE model is very crucial. Li
\cite{6} remarked that instead of Hubble or particle horizons, the
future event horizon should be the IR cutoff which shows
compatibility with the present evolution of the universe. Later on,
it was pointed out \cite{7} that this choice of IR cutoff suffers
the causality problem and proposed the age of the universe as IR
cutoff, called agegraphic DE model. Some other proposals have also
been given for the choice of IR cutoff. Granda and Oliveros \cite{8}
proposed that IR cutoff should be the function of square of the
Hubble parameter and its derivative. This type of IR cutoff is
motivated from the Ricci scalar of the FRW universe \cite{8a}.

The scalar field DE models also belong to the family of dynamical DE
models which explain the DE phenomenon. A wide variety of these
models exists in literature including quintessence, K-essence,
tachyon, phantom, ghost condensates and dilaton \cite{9,10}. Also,
the well-known theories such as the supersymmetric, string and M
theories cannot describe potential of the scalar field
independently. It would be interesting to reconstruct the potential
of DE models so that the scalar fields may describe the cosmological
behavior of the quantum gravity.

The modification in the gravitational part and enhancement of
dimensions in the original general relativity is another way to
handle the DE puzzle. In higher dimensional theories, Kaluza-Klein
(KK) theory \cite{11} has been used extensively for the purpose of
cosmological implications. This theory has been described into ways,
i.e., compact and non-compact forms depending on its fifth
dimension. In its compact form, fifth dimension is like a circle
having very small radius while in non-compact form, it behaves as a
vacuum of 4D geometry. Moreover, the HDE has also been derived in
higher dimensions with the help of the $N$-dimensional mass of the
Schwarzschild black hole \cite{12} known as modified holographic
dark energy (MHDE) model \cite{13,14}. Sharif et al.
\cite{15}-\cite{17} have investigated the evolution as well as
generalized second law of thermodynamics of MHDE with Hubble and
future event horizons as IR cutoffs in the flat and non-flat KK
universe models. In a recent paper \cite{18}, the scalar field
models are constructed for HDE with Hubble horizon and Granda and
Oliveros cutoff as IR cutoff in flat and non-flat universes.

Here we use the Granda and Oliveros \cite{8} IR cutoff for MHDE in
flat KK universe. We discuss the evolution, instability and scalar
field DE models in this scenario. The paper is organized as follows.
Section \textbf{2} contains discussion of the evolution and
instability of MHDE in flat KK universe. In section \textbf{3}, we
reconstruct scalar field models of MHDE. The last section summarizes
the results.

\section{Modified Holographic Dark Energy}

In this section, we make analysis of the equation of state (EoS)
parameter and instability of MHDE in compact flat KK universe
\cite{19} whose metric is given by
\begin{equation}\label{1}
ds^{2}=-dt^{2}+a^{2}(t)[dr^{2}
+r^{2}(d\theta^{2}+\sin\theta^{2}d\phi^{2})+d\psi^{2}],
\end{equation}
where $a(t)$ indicates the cosmic scale factor. The corresponding
field equations are
\begin{eqnarray}\label{2}
H^2&=&\frac{1}{6}\rho_{\Lambda},\\\label{3}
\dot{H}+2H^2&=&-\frac{1}{3} p_{\Lambda},
\end{eqnarray}
where $H$ is the Hubble parameter, dot indicates differentiation
with respect to time and $8\pi G=1$ for the sake of simplicity.
Also, $p_{\Lambda}$ and $\rho_{\Lambda}$ are the pressure and energy
density due to DE respectively. In order to derive MHDE, we use the
formula of the mass of the Schwarzschild black hole in $N$
dimensions
\begin{eqnarray*}
M=\frac{(N-1)A_{N-1}}{16\pi G_{N}}r^{N-2}_{H},
\end{eqnarray*}
where $8\pi G_{N}\equiv M_{*}$ is Planck mass in higher dimensions,
$A_{N-1}$ is the area of $N$ unit spheres and $r_{H}$ represents the
size of the black hole. For KK universe, we take $N=4$ and use the
formula of area, it follows that
\begin{eqnarray*}
M=\frac{3\pi^2 M^3_{*}}{2}r^{5}_{H}.
\end{eqnarray*}
Here, we assume $r_{H}=L=(\mu H^2+\lambda\dot{H})^{-\frac{1}{2}}$
($\mu$ and $\lambda$ are positive constants, this IR cutoff is
proposed by Granda and Oliveros \cite{8}) and $M_{*}$ should be less
than the Planck length. This shows that horizon scale does not make
the mass of five dimensional black hole larger than compactification
scale of KK universe (i.e., of the order of Planck length), while it
reduces its mass. With the help of Cohen et al. \cite{4} relation,
we can derive MHDE in the following form \cite{15}
\begin{equation}\label{4}
\rho_{\Lambda}=3\pi^{2}L^2=\frac{3\pi^{2}}{\mu H^2+\lambda\dot{H}},
\end{equation}
here we take $M_{*}$ to be unity.

The equation of continuity for the MHDE become
\begin{eqnarray}\label{5}
\dot{\rho}_{\Lambda}+4H(\rho_{\Lambda}+p_{\Lambda})=0.
\end{eqnarray}
Equations (\ref{2}) and (\ref{4}) lead to
\begin{equation}\label{6}
\frac{dH^4}{dx}+\frac{4\mu}{\lambda}H^4=\frac{2\pi^2}{\lambda},
\end{equation}
where $x=\ln a$. Solving the above equation, we obtain
\begin{equation}\label{7}
H^4=\frac{\pi^2}{2\mu}+be^{\frac{-4\mu}{\lambda}x},
\end{equation}
where $b$ is an integration constant. The EoS parameter for MHDE can
be obtained by using Eqs.(\ref{4}), (\ref{5}) and (\ref{7})
\begin{eqnarray}\label{8}
\omega_{\Lambda}=-1+\frac{\mu^2
be^{\frac{-4\mu}{\lambda}x}}{\lambda(\pi^2+2\mu
be^{\frac{-4\mu}{\lambda}x})}.
\end{eqnarray}
\begin{figure} \centering
\epsfig{file=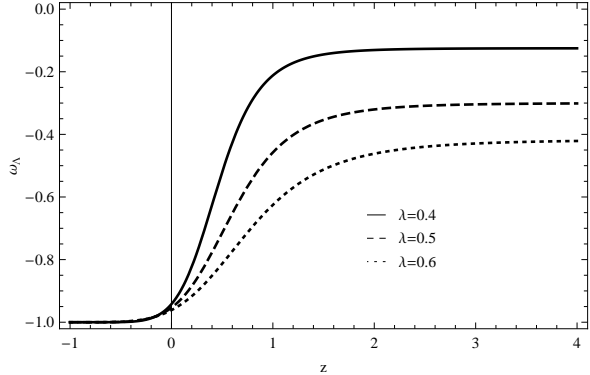,width=.60\linewidth} \caption{Plot of
$\omega_{\Lambda}$ versus $z$ for MHDE.}
\end{figure}
We plot $\omega_{\Lambda}$ versus redshift parameter $z$ by using
$a=a_{0}(1+z)^{-1}$ as shown in Figure $\textbf{1}$ by taking
$\mu=0.7,~b=0.5,~a_{0}=1$ and different values of
$\lambda=0.4,~0.5,~0.6$. It is observed that the EoS parameter
evolutes the universe from quintessence DE era towards vacuum DE
era. This also shows that the parameter $\omega_{\Lambda}$ always
remains in the quintessence region for $\lambda\geq0.5$. However, it
corresponds to the early inflation era for $\lambda<0.5$ in the
early time.

Now we explore the linear perturbation in order to examine the
instability of MHDE. For this purpose, the square of the speed of
sound ($\upsilon^2_s$) is evaluated which characterizes the
stability of DE models. The speed of sound has the form \cite{2a}
\begin{equation}\label{9}
\upsilon_{s}^2=\frac{\dot{p}}{\dot{\rho}}=\frac{p'}{\rho'},
\end{equation}
where prime shows differentiation with respect to $x$. We would like
to mention here that DE models are classically unstable as
$\upsilon^2_s<0$ and vice versa. Equations (\ref{4}), (\ref{5}),
(\ref{7}) and (\ref{9}) yield
\begin{eqnarray}\label{10}
\upsilon^2_{s}&=&\frac{\pi^2(\mu-\lambda)-\mu
b(2\lambda-\mu)e^{\frac{-4\mu}{\lambda}x}}{\lambda(\pi^2+2\mu
be^{\frac{-4\mu}{\lambda}x})}.
\end{eqnarray}
We plot the speed of sound versus $z$ as shown in Figure \textbf{2}
by keeping the same values of constant parameters as given above.
This shows that the MHDE remains unstable in the early epoch and
stable for the present and later time.
\begin{figure} \centering
\epsfig{file=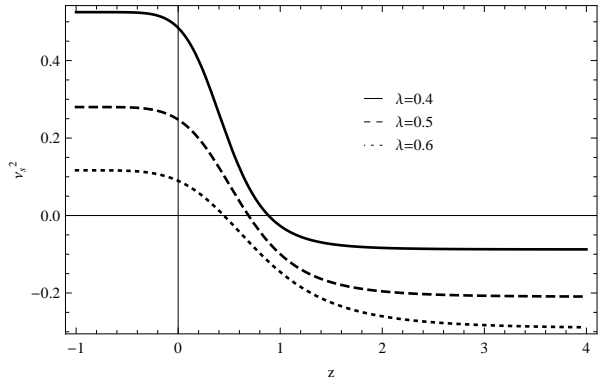,width=.60\linewidth} \caption{Plot of
$\upsilon_{s}^2$ versus $z$ for MHDE.}
\end{figure}

\section{Correspondence of MHDE with Scalar Field Models}

The current prediction of accelerated expansion of the universe has
roots in the early inflationary epoch. It is argued that the
inflation field has similar properties as the cosmological constant.
Moreover, inspired by the inflation field, scalar field models have
been proposed. Scalar field models are used in explaining the DE
phenomenon due to their dynamical nature. These models describe the
quintessence behavior of the universe and also provide the effective
description of DE models through reconstruction scenario. In our
case, the EoS for MHDE obeys this argument and hence it would be
interesting to reconstruct the scalar field DE models in this
scenario. We discuss all the results graphically with the same
values of the constant parameters as given in the previous section.

\subsection{Quintessence Dark Energy Model}

This model is originated in the light of scalar field $\phi$ which
is minimally coupled with gravity. The energy density and pressure
of the quintessence DE model are given by \cite{10}
\begin{eqnarray}\label{11}
\rho_{q}=\frac{1}{2}\dot{\phi}^{2}+V(\phi),\quad
p_{q}=\frac{1}{2}\dot{\phi}^{2}-V(\phi),
\end{eqnarray}
where $\dot{\phi}^{2}$ is the kinetic energy and $V(\phi)$ is the
potential of scalar field. This equation gives the EoS parameter as
\begin{equation*}
\omega_{q}=\frac{\dot{\phi}^{2}-2V(\phi)}{\dot{\phi}^{2}+2V(\phi)}.
\end{equation*}
We identify $\rho_{q}=\rho_{\Lambda}$ and $p_{q}=p_{\Lambda}$ to
establish the correspondence between MHDE and quintessence scalar
field. Thus, it follows from Eq.(\ref{11}) that
\begin{eqnarray}\label{12}
\dot{\phi}^{2}&=&\frac{\mu\sqrt{6b}e^{\frac{-2\mu}{\lambda}x}}{\sqrt{\lambda(\pi^2+2\mu
be^{\frac{-4\mu}{\lambda}x})}},\\\label{13}
V(\phi)&=&\frac{6\pi^2\lambda-3b\mu(\mu-4\lambda)e^{\frac{-4\mu}{\lambda}x}}
{\lambda\sqrt{2\mu(\pi^2+2\mu ce^{\frac{-4\mu}{\lambda}x})}}.
\end{eqnarray}
The value of scalar field turns out to be
\begin{eqnarray}\label{14}
\phi=\int^{x}_{0}\sqrt{\frac{6\mu^2be^{\frac{-4\mu}{\lambda}x}}{\lambda(\pi^2+2\mu
be^{\frac{-4\mu}{\lambda}x})}}dx.
\end{eqnarray}
The evolutionary trajectories of the scalar field $\phi$ and the
corresponding scalar potential are shown in Figures \textbf{3} and
\textbf{4}. Here we use the initial condition of the scalar field
$\phi(0)=0$. We notice that the quintessence field increases
initially but becomes flat at high redshift. This shows that the
field decreases gradually with the expansion of the universe. Also,
we observe that the quintessence potential becomes more steeper with
the decrease of $\lambda$ and tends to flat in the later epoch.
\begin{figure} \centering
\epsfig{file=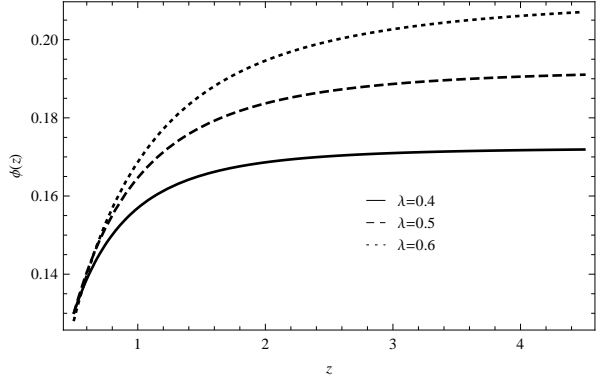,width=.60\linewidth} \caption{Plot of $\phi(z)$
versus $z$ for quintessence model.}
\end{figure}
\begin{figure} \centering
\epsfig{file=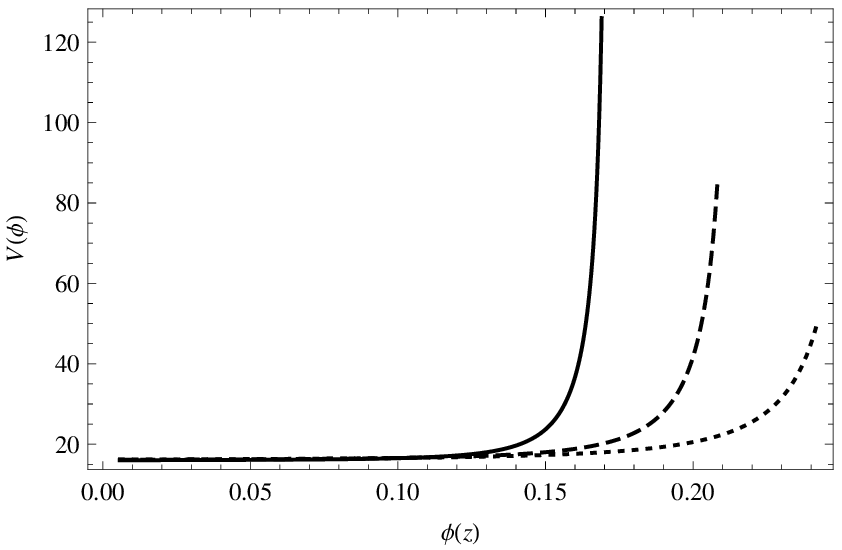,width=.60\linewidth} \caption{Plot of $V(\phi)$
versus $\phi(z)$ for quintessence model.}
\end{figure}

\subsection{Tachyon Dark Energy Model}

The energy and pressure of the tachyon DE model has the form
\cite{10}
\begin{eqnarray}\label{15}
\rho_{t}=\frac{V(\phi)}{\sqrt{1-\dot{\phi}^{2}}},\quad
p_{t}=-V(\phi)\sqrt{1-\dot{\phi}^{2}}.
\end{eqnarray}
The corresponding EoS parameter is
\begin{equation}\label{16}
\omega_{t}=\dot{\phi}^{2}-1.
\end{equation}
The correspondence between MHDE and tachyon model is obtained for
$\rho_{t}=\rho_{\Lambda}$ and $p_{t}=p_{\Lambda}$, which leads to
\begin{eqnarray}\label{17}
\dot{\phi}^{2}&=&\frac{\mu^2be^{\frac{-4\mu}{\lambda}x}}{\lambda(\pi^2+2\mu
ce^{\frac{-4\mu}{\lambda}x})},\\\label{18}
V(\phi)&=&\sqrt{\frac{18(\pi^2\lambda+\mu
b(2\lambda-\mu)e^{\frac{-4\mu}{\lambda}x})}{\mu\lambda}}.
\end{eqnarray}
Equations (\ref{7}) and (\ref{17}) give
\begin{eqnarray}\label{19}
\phi'(x)=\left(\frac{2\mu}{\pi^2+2\mu
ce^{\frac{-4\mu}{\lambda}x}}\right)^{\frac{1}{4}}
\sqrt{\frac{\mu^2be^{\frac{-4\mu}{\lambda}x}}{\lambda(\pi^2+2\mu
be^{\frac{-4\mu}{\lambda}x})}}.
\end{eqnarray}
We solve and plot it against $z$ as shown in Figure $\textbf{5}$.
The evolution of tachyon field is very much similar to the
quintessence field. The tachyon field attains the maximum value at
early epoch and then decreases and goes towards zero at the present
epoch. Also, we remark that the tachyon field rolls down potential
slowly with the expansion of the universe as shown in Figure
\textbf{6}.
\begin{figure} \centering
\epsfig{file=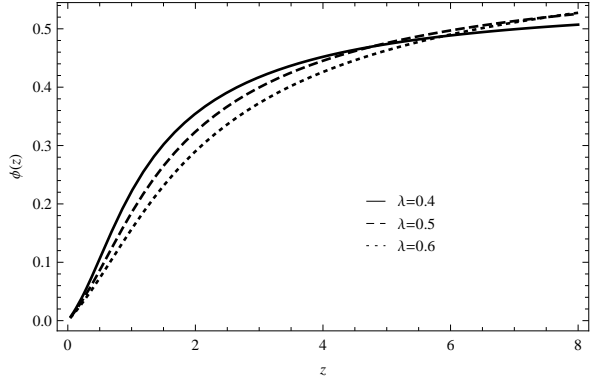,width=.60\linewidth} \caption{Plot of $\phi(z)$
versus $z$ for tachyon model.}
\end{figure}
\begin{figure} \centering
\epsfig{file=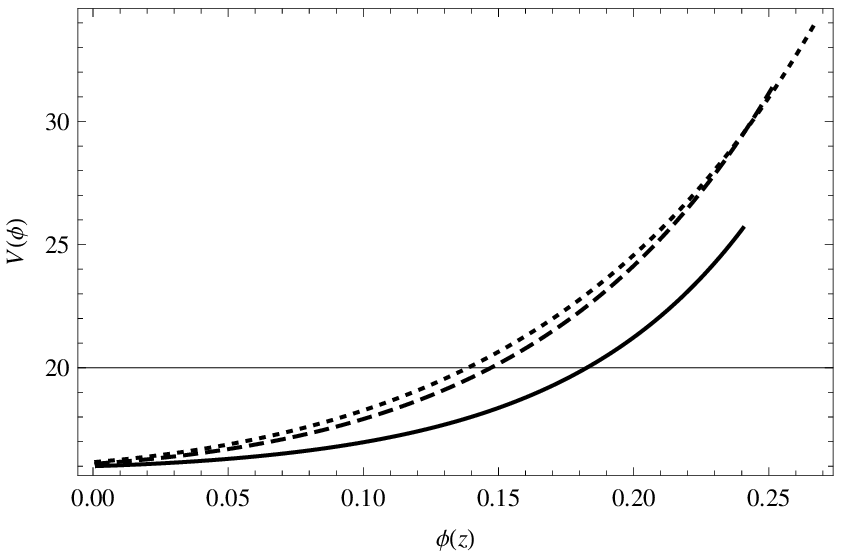,width=.60\linewidth} \caption{Plot of $V(\phi)$
versus $\phi(z)$ for tachyon model.}
\end{figure}

\subsection{K-essence Dark Energy Model}

The energy density and pressure of this model are \cite{10}
\begin{eqnarray}\label{20}
\rho_{k}=V(\phi)(-\chi+3\chi^{2}),\quad
p_{k}=V(\phi)(-\chi+\chi^{2}),
\end{eqnarray}
where $\chi=\frac{1}{2}~\dot{\phi}^2$ and $V(\phi)$ represents the
scalar potential of K-essence model. This leads to the following EoS
parameter for tachyon DE model
\begin{equation}\label{21}
\omega_{k}=\frac{1-\chi}{1-3\chi}.
\end{equation}
\begin{figure} \centering
\epsfig{file=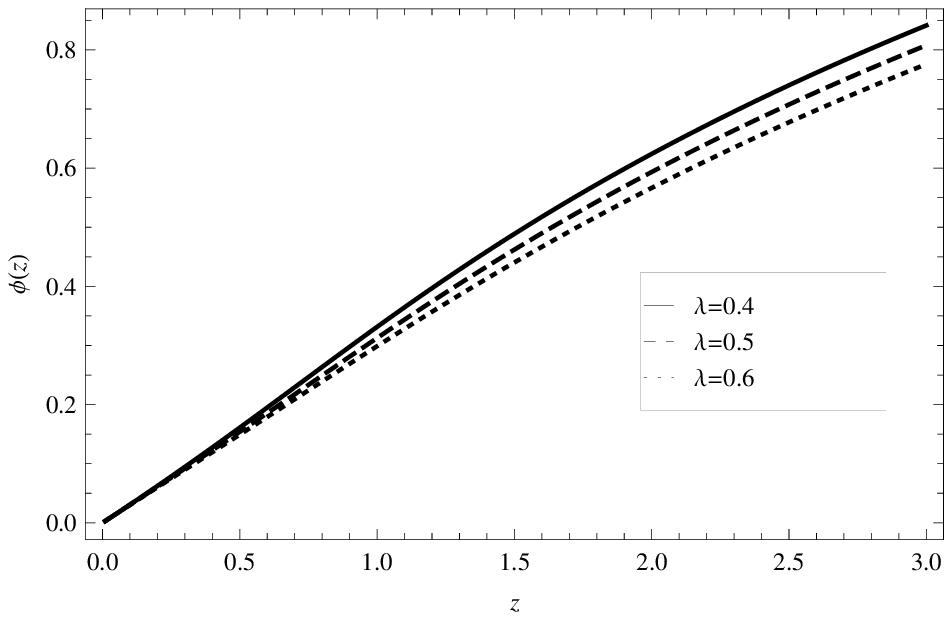,width=.60\linewidth} \caption{Plot of $\phi(z)$
versus $z$ for K-essence model.}
\end{figure}
\begin{figure} \centering
\epsfig{file=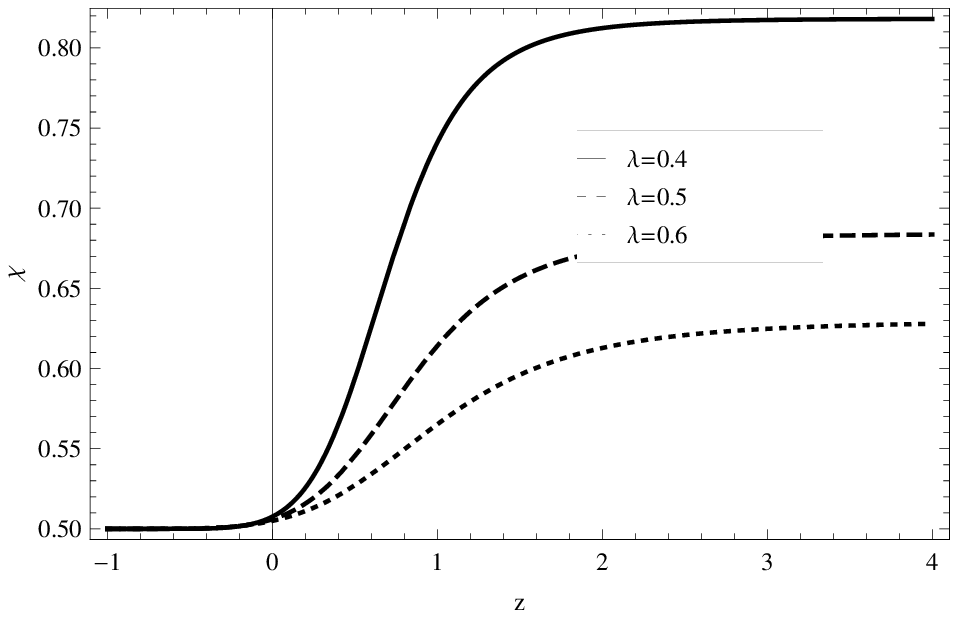,width=.60\linewidth} \caption{Plot of $\chi$
versus $z$ for K-essence model.}
\end{figure}
\begin{figure} \centering
\epsfig{file=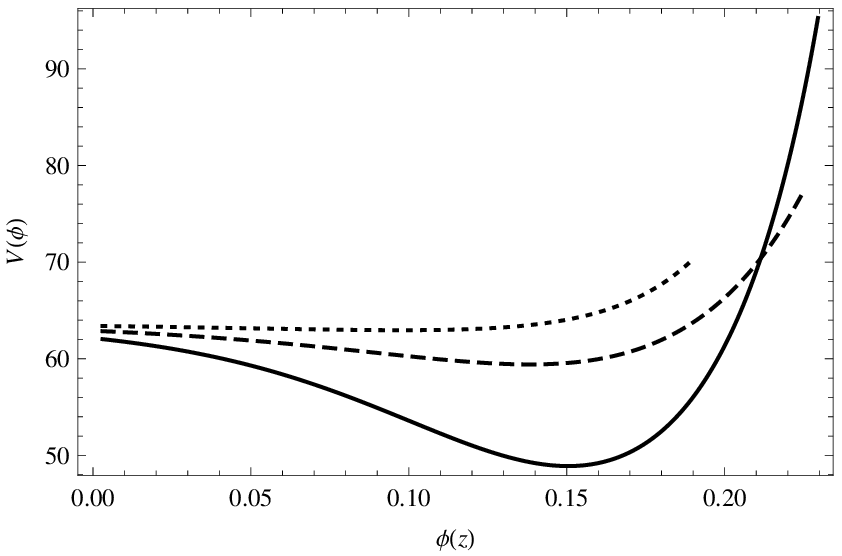,width=.60\linewidth} \caption{Plot of $V(\phi)$
versus $\phi(z)$ for K-essence model.}
\end{figure}
We set $\rho_{k}=\rho_{\Lambda}$ and $p_{k}=p_{\Lambda}$ for the
correspondence between MHDE and K-essence model and obtain
\begin{eqnarray}\label{22}
\chi&=&\frac{2\pi^2\lambda+\mu
b(4\lambda-\mu)e^{\frac{-4\mu}{\lambda}x}}{4\pi^2\lambda+\mu
b(8\lambda-3\mu)e^{\frac{-4\mu}{\lambda}x}},\\\label{23}
V(\phi)&=&\frac{3(1-3\omega_{\Lambda})^2}{1-\omega_{\Lambda}}\left(\frac{2\mu}{\pi^2+2\mu
ce^{\frac{-4\mu}{\lambda}x}}\right)^{\frac{1}{2}}.
\end{eqnarray}

Finally, the expression $\chi=\frac{1}{2}\dot{\phi}^2$ leads to
\begin{eqnarray}\label{24}
\phi'(x)=\left(\frac{2\mu}{\pi^2+2\mu
ce^{\frac{-4\mu}{\lambda}x}}\right)^{\frac{1}{4}}\sqrt{\frac{2(2\pi^2\lambda+\mu
b(4\lambda-\mu)e^{\frac{-4\mu}{\lambda}x})}{4\pi^2\lambda+\mu
b(8\lambda-3\mu)e^{\frac{-4\mu}{\lambda}x}}}.
\end{eqnarray}
Figure $\textbf{7}$ shows that the K-essence scalar field decreases
with the increment of MHDE parameter $\lambda$. The EoS of tachyon
DE model is compatible with the accelerated expansion of the
universe in the range $\frac{1}{3}<\omega_{\Lambda}<\frac{2}{3}$. It
is noted that the kinetic term is consistent with this EoS for
$\lambda=0.5,~0.6$, but it is inconsistent for $\lambda\leq0.4$ as
shown in Figure $\textbf{8}$. Also, the plot $\textbf{9}$ indicates
that the K-essence potential increases with the increase of the
field $\phi$. It rolls down the potential because K-essence scalar
field decreases with the expansion of the universe.

\subsection{Dilaton Dark Energy Model}

The Lagrangian of dilaton field describes the pressure of scalar
field given by \cite{9}
\begin{equation}\label{25}
p_d=-\chi+me^{n\phi}\chi^2,
\end{equation}
where $m$ and $n$ are taken as positive constants. The corresponding
energy density is
\begin{eqnarray}\label{26}
\rho_{d}&=&-\chi+3me^{n\phi}\chi^{2}.
\end{eqnarray}
This model has the EoS parameter
\begin{equation}\label{27}
\omega_{d}=\frac{-1+me^{n\phi}\chi}{-1+3m~e^{n\phi}\chi}.
\end{equation}
The replacement of $\rho_{d}$ by $\rho_{\Lambda}$ and $p_{d}$ by
$p_{\Lambda}$ (for correspondence) gives
\begin{eqnarray}\label{28}
e^{n\phi}\chi&=&\frac{2\pi^2\lambda+\mu
b(4\lambda-\mu)e^{\frac{-4\mu}{\lambda}x}}{m(4\pi^2\lambda+\mu
b(8\lambda-3\mu)e^{\frac{-4\mu}{\lambda}x})}.
\end{eqnarray}
Its plot against $z$ with $m=1.5$ and $n=0.05$ is shown in Figure
\textbf{10}. We observe that $e^{n\phi}\chi $ shows consistency with
the expanding scenario of the universe predicated by EoS of this
model. The solution of the above equation follows
\begin{eqnarray}\nonumber
\phi(x)&=&\frac{2}{n}\ln\left[1+\frac{n}{\sqrt{2m}}\int^{x}_{0}
\sqrt{\frac{2\pi^2\lambda+\mu
b(4\lambda-\mu)e^{\frac{-4\mu}{\lambda}x}}{4\pi^2\lambda+\mu
b(8\lambda-3\mu)e^{\frac{-4\mu}{\lambda}x}}}\right.\\\label{29}
&\times&\left.\left(\frac{2\mu}{\pi^2+2\mu
be^{\frac{-4\mu}{\lambda}x}}\right)^{\frac{1}{4}}dx\right].
\end{eqnarray}
Its graphical presentation is shown in Figure \textbf{11} which
exhibits direct proportionality with respect to $z$ leading to the
scaling solutions for dilaton model \cite{9}.
\begin{figure} \centering
\epsfig{file=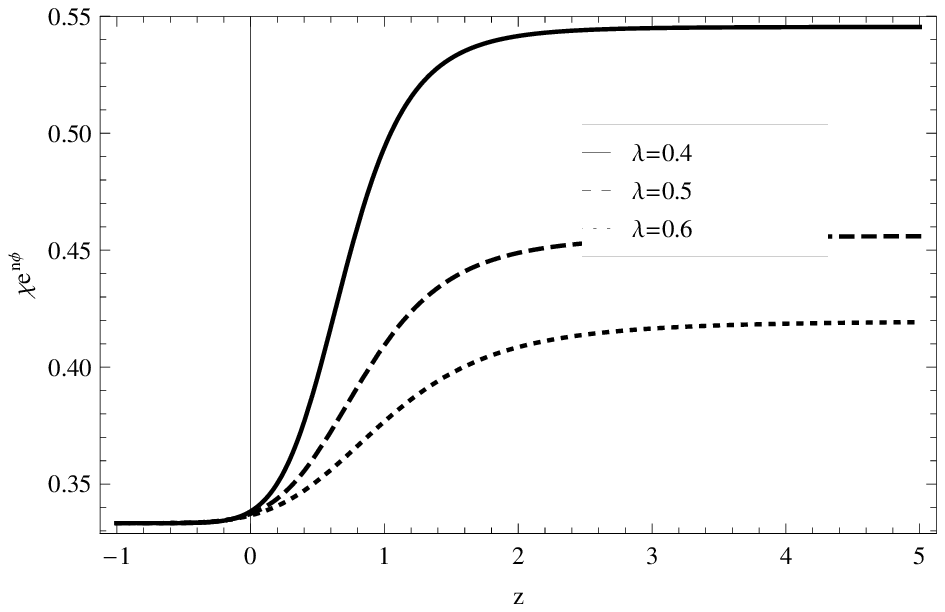,width=.60\linewidth} \caption{Plot of
$e^{n\phi}\chi$ versus $z$ for dilaton field.}
\end{figure}
\begin{figure} \centering
\epsfig{file=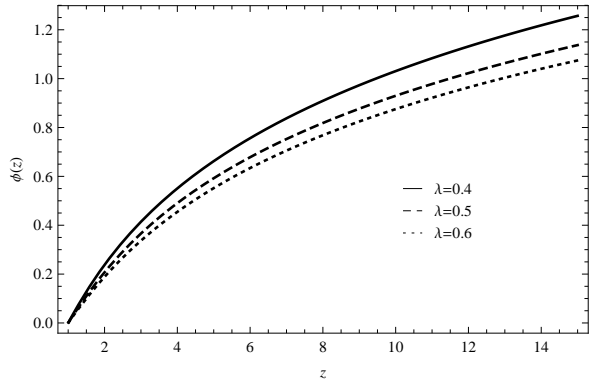,width=.60\linewidth} \caption{Plot of $\phi(z)$
versus $z$ for dilaton field.}
\end{figure}

\section{Summary}

It is believed that our universe expands with accelerated expansion
and the scalar field DE models act as an effective theories of an
underlying phenomenon of DE. We investigate the evolution of MHDE,
instability and reconstructed scalar field DE models. We consider
MHDE with IR cutoff as a function of the Hubble parameter and its
derivative in the flat Kaluza-Klein universe. This type of DE
density avoids the causality problem due to dependence on the local
quantities. The MHDE parameter $\lambda$ plays the crucial role in
evaluating the results. We have used the best fit values of
$\mu=0.7$ and $\lambda=0.4,~0.5,~0.6$ obtained by Wang and Xu
\cite{21} through type Ia supernovae, baryon acoustic oscillations,
CMBR and the observational Hubble data. The results are summarized
as follows.
\begin{itemize}
\item The EoS parameter $\omega_{\Lambda}$ shows
transition from quintessence DE era ($-\frac{1}{3}<\omega<-1$)
towards vacuum DE era ($\omega=-1$) as displayed in Figure
$\textbf{1}$. This type of behavior has led to reconstruct the
scalar field DE models. Also, our result of the present value of
$\omega_{\Lambda}=-0.95$ at $z=0$ (for $\lambda=0.4,~0.5,~0.6$) is
almost closer to the value obtained for HDE in GR \cite{6,8,22,23}.
\item It is interesting to check the viability of the MHDE model due
to its dependence upon local quantities. Using the squared speed of
sound, we have found that the MHDE with new IR cutoff is unstable
for early epoch but stable for the later time as shown in Figure
$\textbf{2}$. It is mentioned here that the MHDE is always stable
for $\lambda<0.35$, while in GR, the new HDE is unstable \cite{24}.
In addition, the Chaplygin and tachyon Chaplygin gases are stable as
$\upsilon^2_s>0$, but the holographic \cite{25}, agegraphic
\cite{26} and QCD ghost DE \cite{27} models are classically unstable
as $\upsilon^2_s<0$.
\item Finally, we have provided the correspondence of MHDE with
scalar field DE models including quintessence, tachyon, K-essence
and dilaton models. In all these models, the scalar field shows the
decreasing behavior with the expansion of the universe. The scalar
potential increases and becomes steeper with the increase of scalar
field $\phi$ for quintessence, tachyon and K-essence DE models. The
plots of scalar field DE models are given in Figures
\textbf{3}$-$\textbf{11}. In K-essence and dilaton DE models, the
kinetic terms exactly lie in the required region (where EoS
parameter predicts the accelerated expansion of the universe). We
would like to remark here that the results of scalar field and
corresponding potential are consistent with the current status of
the universe. These are also consistent with the results of Zhang
\cite{22} for quintessence HDE, for tachyon HDE \cite{23,28,29} and
Rozas-Fern$\acute{a}$ndez \cite{30} for dilaton HDE models.
\end{itemize}

\end{document}